\documentclass[aps,prl,notitlepage,onecolumn,12pt]{revtex4-1}

\usepackage{amsmath}
\usepackage{amssymb}
\usepackage{graphicx,subcaption}
\usepackage{ulem}

\newcommand\be{\begin{equation}}
\newcommand\ba{\begin{eqnarray}}
\newcommand\ee{\end{equation}}
\newcommand\ea{\end{eqnarray}}

\newcommand{\dd}[1]{\mathrm{d}#1\,}

\begin{document}
\title {On the Classical and Quantum Stability of a Cosmic Ghost}
\author{Stephon Alexander}
\affiliation{Department of Physics, Brown University, Providence, RI 02906}
\author{Sam Cormack}
\affiliation{Department of Physics and Astronomy, Dartmouth College}
\author{David Lowe}
\affiliation{Department of Physics, Brown University, Providence, RI 02906}
\author{Robert Sims}
\affiliation{Department of Physics and Astronomy, Dartmouth College}

%Hanover, NH 03755} 
\begin{abstract}
Ghost fields have reemerged in a handful of phenomenologically motivated cosmological and particle physics scenarios, and most recently in a cyclic mechanism to address the fine-tuning of gauge couplings in the standard model.  We study the classical and quantum stability of a ghost-dilaton system coupled to a gauge sector and find that this system is classically stable due to the existence of limit cycles in phase space.  We also analyze the coupled gauge invariant classical perturbations and find a range of phenomenologically viable parameters where the system is stable.  We also discuss ways to avoid both quantum and vacuum instabilities by either having a ghost condensate or Classicon configurations.
\end{abstract}

\date{\today}

\maketitle

%%%%%%%%%%%%%%%%%%%%%%%%%%%%%%%%%%%%%%%%%%%%%%%%%%%%%%%%%%%%%%%%%%%%%%%%%%%%%%%%%%%%%%%%%%%%%%
\section{Introduction}

The issue of ghost fields and their stability have been a subject of interest over the past years \cite{Mukhanov:1990me,ArkaniHamed:2003uy,Krotov:2004if,Adams:2006sv,Muk,Barrow:2009sj,Dvali:2012zc}.  The common lore is that fields with negative kinetic energy could have an unbounded vacuum decay into photons and gravitons which place tight bounds on their existence\cite{Trodden,Cline}.  However, ghosts have been revisited in the context of infrared modifications to gravity to address a range of phenomenological issues, such as dark energy, cosmic inflation, a probe of Lorentz violation, and renormalizable deformation of perturbative quantum gravity.

Ghost fields seem to be inevitable in bouncing and cyclic models such as Ekpyrotic, Anamorphic and Matter-Bounce   \cite{Steinhardt:2001st,Steinhardt:2001vw,Novello:2008ra,Cai:2012va,Lehners:2008vx,Brandenberger:2012zb}.  This is because the Friedmann equations require the Hubble parameter to go to zero at the bounce and a negative energy contribution in the Hamiltonian is needed to accomplish this requirement.  Recently, a cyclic cosmology model was proposed to deal with the fine tuning issues in gauge couplings in the standard model.  Instead of using the inflationary multiverse idea to populate pocket universes with different couplings, the authors demonstrate that in a cyclic universe, the gauge couplings can undergo a quasi-random walk during the bounce and stabilize during the expanding phase.  Therefore, we happen to live in an expansion epoch where the couplings have dynamically evolved to be compatible with the measured values in our standard model.  

This ``cyclic-multiverse'' model \cite{Alexander:2015pda} implements a dilaton-like field which naturally acts as coupling constants for gauge theories.  It is remarkable that in this model, the dilaton field plays a dual role as the agent that leads to the bounce and the coupling constant.  Despite its promise, issues of stability arise since the dilaton field looks like a ghost.   However, because of the non-linear couplings and time-dependence, it is important to address an explicit stability analysis for this theory in particular. 

In this work we will derive the gauge invariant cosmological perturbations of the ghost-dilaton-gauge system and perform a classical, linear stability analysis.  We then discuss the classical nonlinear vacuum stability issues particular to this model, extensions to ghost condensate models and conclude with a future research directions regarding UV completion and non-perturbative physics.

\section{Background Equations}
We are investigating the stability of a ghost scalar field with a periodic potential with a dilatonic coupling to a $U(1)$ gauge field. The action for our model is
\begin{equation}
\label{eq:action}
S =\int\dd{^4x}\sqrt{-g}\left[\frac{R}{16\pi G}+\frac{1}{2}\partial_\mu \psi\partial^\mu\psi-V(\psi) -\frac{1}{4}g_0 e^{-2\psi/M_p} F_{\mu\nu}F^{\mu\nu}
\right],
\end{equation}
where with our sign convention $(-,+,+,+)$ the kinetic term for the field $\psi$ has the wrong sign. We use a negative periodic potential
\begin{equation}
V(\psi) = -\Lambda^4(1+\cos(\psi/f)).
\end{equation}
First, we will investigate the classical, linear stability. At the background level, we assume an FRW metric in conformal coordinates,
\begin{equation}
ds^2 = a^2(\eta)\left(-d\eta^2+ \gamma_{ij}dx^idx^j\right)
\end{equation}
where $\gamma_{ij} = \delta_{ij}\left[1+\frac{1}{4}\mathcal{K}\left(x^2+y^2+z^2\right)\right]^{-2}$ is the spatial metric and $\mathcal{K}$ the spatial curvature. Let us consider the equation of motion for the homogeneous background part of the ghost field $\psi=\psi(\eta)$,
\begin{equation}
\label{eq:psiEOM}
\psi'' + 2\mathcal{H}\psi' - a^2 \frac{\partial V}{\partial\psi} +a^2 \frac{g_0}{2M_p}e^{-2\psi/M_p}F_{\mu\nu}F^{\mu\nu}=0
\end{equation}
where a prime denotes a derivative with respect to conformal time.

\begin{figure}
\includegraphics[scale=1]{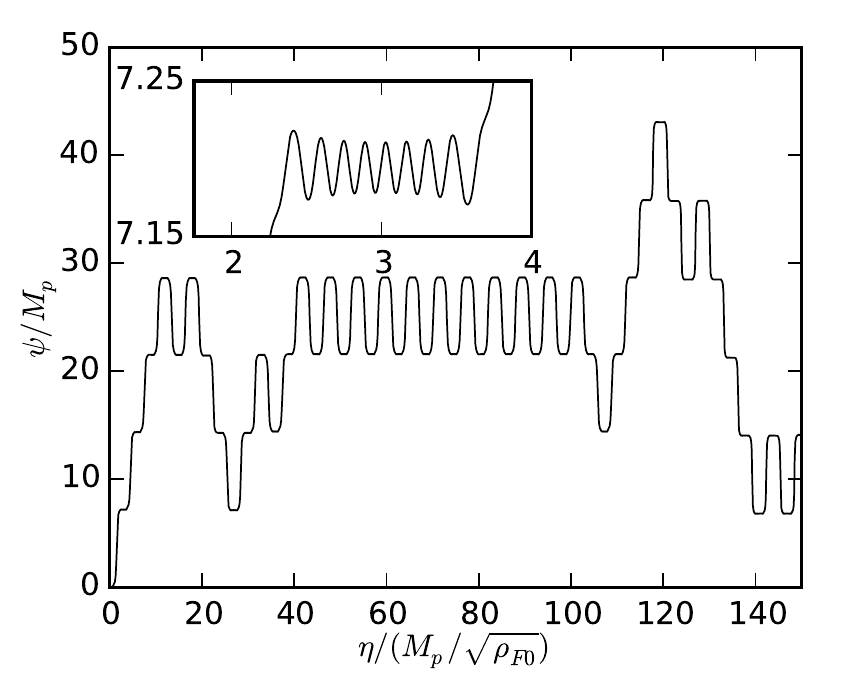}
\caption{\label{fig:psisol}Example of the behaviour of the background solution for the ghost field $\psi$ as a function of conformal time. The inset shows the behaviour between two bounces.  The parameters used in this solution are $f=10^{-2}M_p$ and $\Lambda^4=\rho_{F0}$.}
\end{figure}

To solve for the evolution of the homogeneous background quantities, we assume that the gauge field is in the form of radiation. In that case $F_{\mu\nu}F^{\mu\nu}=0$, and we can ignore the last term in equation \eqref{eq:psiEOM}. The Friedmann equations for the evolution of the spacetime are
\begin{gather}
\label{eq:Fried1}
\mathcal{H}^2 = \frac{1}{3M_p^2}\left[-\frac{\psi'^2}{2}-a^2\Lambda^4\left(1+\cos(\psi/f)\right)+\frac{\rho_{F0}}{a^2}\right]-\mathcal{K}\\
\label{eq:Fried2}
\mathcal{H}' = \frac{1}{3M_p^2}\left[\psi'^2-a^2\Lambda^4\left(1+\cos(\psi/f)\right)-\frac{\rho_{F0}}{a^2}\right]
\end{gather}
where $\mathcal{H}\equiv a'/a$ and $\rho_{F0}$ is the radiation energy density at $a=1$. In order to find cyclic solutions, we take the curvature to be positive. In figure \ref{fig:psisol} we show an example of the behaviour of the field $\psi(\eta)$. During this evolution, the scale factor oscillates with the time scale at its minimum very short compared to rest of the evolution. We refer to this period, where the scale factor goes from decreasing to increasing, as a bounce. During each bounce the field $\psi$ jumps quickly from one value to another. Between bounces, $\psi$ exhibits linearly stable oscillations around the maximum of its potential, since the kinetic term is of the wrong sign. These oscillations can be seen in the inset of figure \ref{fig:psisol}. The changing value of $\psi$ from one cycle to another changes the effective coupling constant of the gauge field via the dilatonic coupling in equation \eqref{eq:action}.

\begin{figure}
\begin{subfigure}{0.47\textwidth}
\includegraphics[scale=0.93]{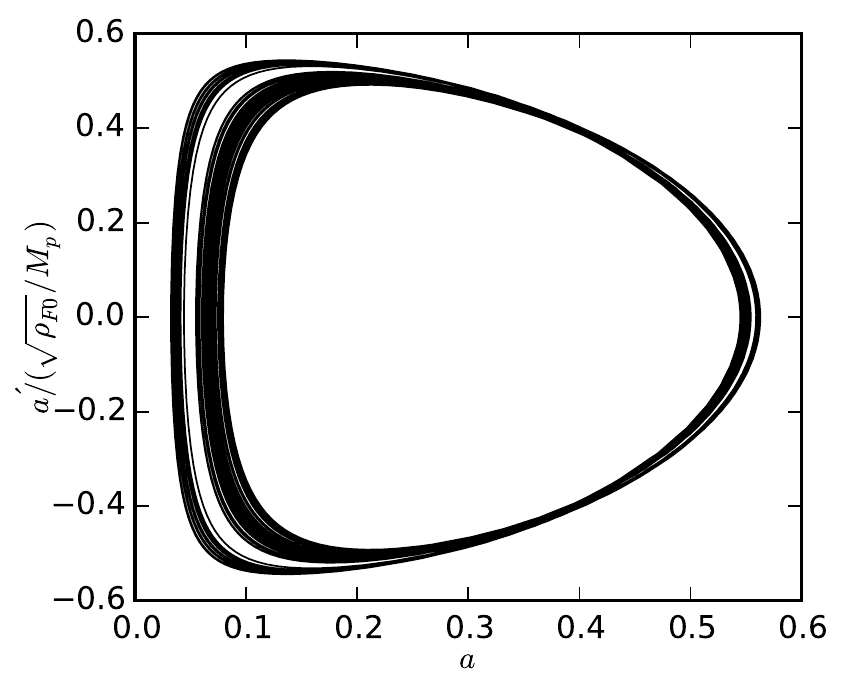}
\caption{Phase portrait showing limit cycle for $f =10^{-2}M_p$}
\label{fig:phaseportE-2}
\end{subfigure}
\begin{subfigure}{0.47\textwidth}
\includegraphics[scale=0.93]{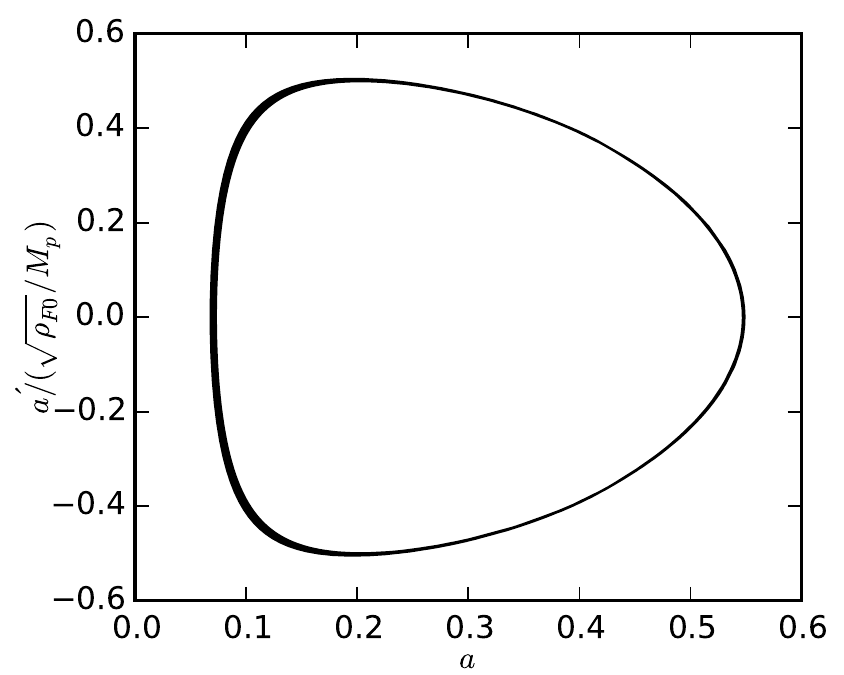}
\caption{Phase portrait showing limit cycle for $f =10^{-3}M_p$}
\label{fig:phaseportE-3}
\end{subfigure}
\begin{subfigure}{0.47\textwidth}
\includegraphics[scale=0.93]{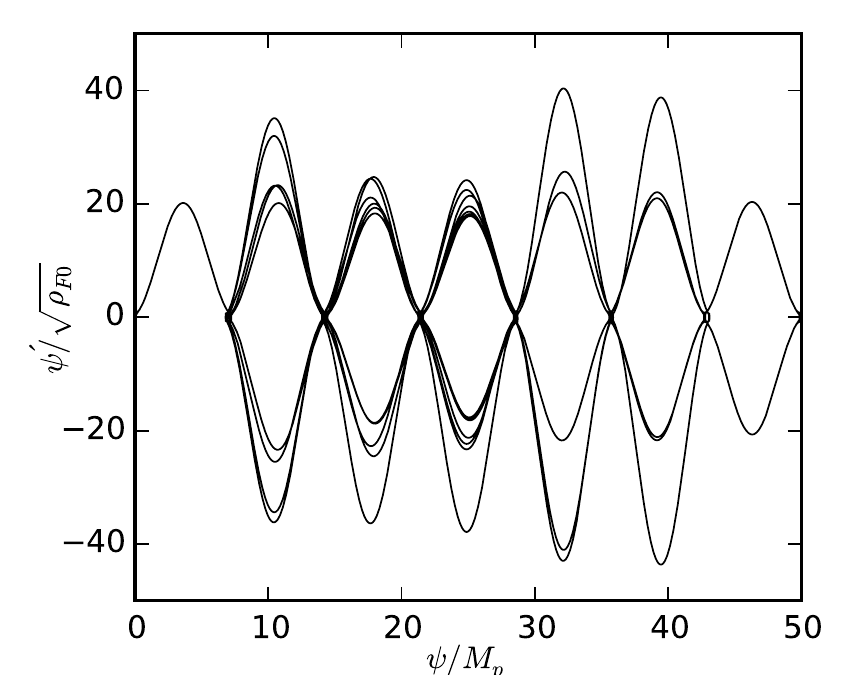}
\caption{Phase portrait for ghost field, $f=10^{-2}M_p$.}
\label{fig:phaseportE-4}
\end{subfigure}
\begin{subfigure}{0.47\textwidth}
\includegraphics[scale=0.93]{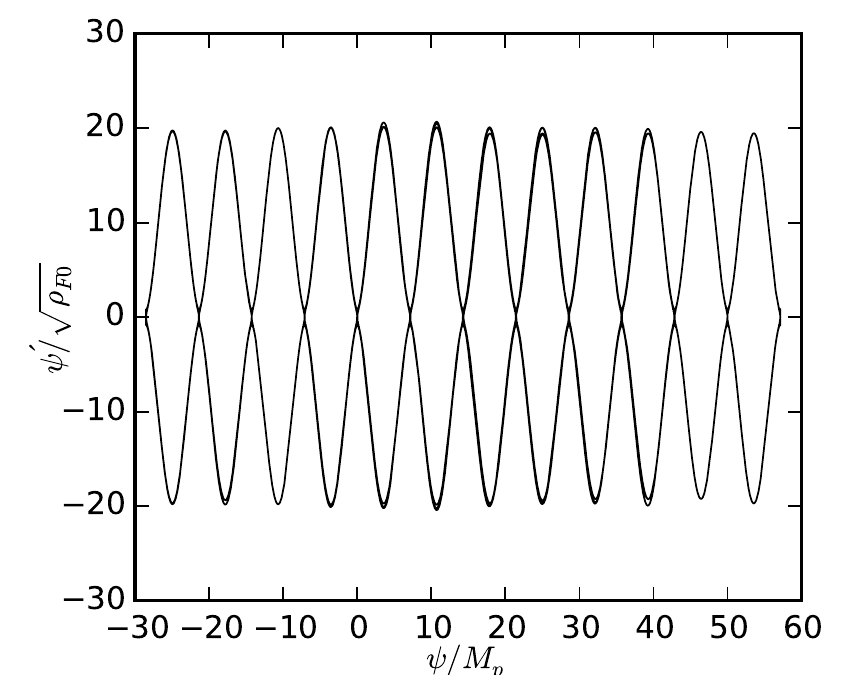}
\caption{Phase portrait for ghost field, $f=10^{-3}M_p$.}
\label{fig:phaseportE-5}
\end{subfigure}
\end{figure}

We can study the stability of the background solutions by analyzing phase portraits of the scale factor. These are plotted in figures \ref{fig:phaseportE-2} and \ref{fig:phaseportE-3} for $f=10^{-2}M_p$ and $f=10^{-3}M_p$ respectively. We see that the scale factor solution is almost periodic. The precise evolution differs from cycle to cycle, but the overall solution remains confined to a band, indicating the stability of the solutions. We can also see that for the smaller value of $f$, the band of solutions is much tighter, showing that the solutions become closer to periodic for smaller values of $f$.

The phase portraits for the ghost field $\psi$, figures \ref{fig:phaseportE-4} and \ref{fig:phaseportE-5}, show the overall behaviour of the ghost field. During the expansion and contraction, the evolution of the field is confined to the intersection near the $\psi'=0$ axis. It is only during the bounce phases where the large jumps occur which take the ghost field to different maxima of the potential. As with the scale factor, the evolution of the ghost field becomes more tightly confined to cycles as $f$ becomes smaller.

To investigate linear instabilities caused by inhomogeneous perturbations, we perturb both the ghost field and the metric. For now, we ignore the gauge field.  The scalar perturbation of the background Friedmann-Robertson-Walker metric in the Newtonian gauge gives
\begin{equation}
ds^2 = a^2\left(\eta\right)\left(-\left(1+2\phi\right)d\eta^2 + \left(1-2\theta\right)\gamma_{ij}dx^idx^j\right)
\end{equation}
By perturbing the ghost field, $\psi \mapsto \psi(\eta) + \chi(\eta,\mathbf{x})$, we obtain the Einstein equation $G^\mu_{\; \nu} = M_p^{-2} T^\mu_{\; \nu}$ to first order in perturbations, $\phi$, $\theta$ and $\chi$. For a scalar field source, there is no anisotropic stress, and we have $\phi=\theta$. The field equations for the metric perturbation $\phi$ are then
\begin{gather}
\label{eq:PertEE00}
\nabla^2 \phi - 3\mathcal{H}\phi' - 3\phi \left(\mathcal{H}^2 - \mathcal{K}\right) = \frac{1}{2M_p^2} \left(\phi \left(\psi'\right)^2 -\psi'\chi' + a^2 \frac{\partial V}{\partial \psi}\chi\right) \\
\label{eq:PertEE0i}
\mathcal{H}\phi + \phi' = -\frac{1}{2M_p^2}\psi'\chi \\
\label{eq:PertEEij}
\phi'' + 3\mathcal{H}\phi'+\left(2\mathcal{H}'+\mathcal{H}^2 - \mathcal{K}\right)\phi = \frac{1}{2M_p^2}\left(\phi\left(\psi'\right)- \psi'\chi' - a^2\frac{\partial V}{\partial \psi}\chi\right).
\end{gather}
These are the same as the standard equations for scalar perturbations sourced by a scalar field except that those terms derived from the kinetic term of the ghost field have the opposite sign \cite{Mukhanov:1990me}. With the second order Lagrangian, the equation of motion for the perturbation of the ghost field can be found.  This equation of motion, along with a combination of equations \eqref{eq:PertEE00} and \eqref{eq:PertEEij}, give wave equations for the two perturbative fields
\begin{gather}
\chi'' - \nabla^2\chi + 2\mathcal{H}\chi' - a^2 \frac{\partial^2 V}{\partial \psi^2}\chi = 2\phi\psi'' + 4\psi'\left(\mathcal{H}\phi+\phi'\right) \\
\phi'' - \nabla^2\phi + 6\mathcal{H}\phi' + \left(2\mathcal{H}' + 4\mathcal{H}^2 - 4\mathcal{K}\right)\phi = - \frac{a^2}{M_p^2} \frac{\partial V}{\partial \psi}\chi
\end{gather}
We can simplify the coupling terms in the equations by using the background equation of motion for $\psi$ and equation \eqref{eq:PertEE0i}. We again assume that the gauge field just contributes a radiation density $\rho_F$ to the background evolution. It will then have no direct effect on the background equation for $\psi$ since we will have $F_{\mu\nu}F^{\mu\nu}=0$. The equation of motion for $\psi$ is then
\begin{equation}
\label{eq:psiEOM2}
\psi'' + 2\mathcal{H}\psi' -a^2\frac{\Lambda^4}{f}\sin\left(\frac{\psi}{f}\right) = 0.
\end{equation}
In Fourier space, the equations for the perturbations become
\begin{gather}
\label{eq:chi_k}
\chi_k'' + 2\mathcal{H}\chi_k' + \left(k^2+\frac{2}{M_p^2}\psi'^2-a^2\frac{\Lambda^4}{f^2}\cos(\psi/f)\right)\chi_k = 2\psi''\phi_k\\
\label{eq:phi_k}
\phi_k'' + 2\mathcal{H}\phi_k' + \left(k^2+2\mathcal{H}'-4\mathcal{K}\right)\phi_k = -\frac{1}{M_p^2}\psi''\chi_k.
\end{gather}

\begin{figure}
\includegraphics[scale=1]{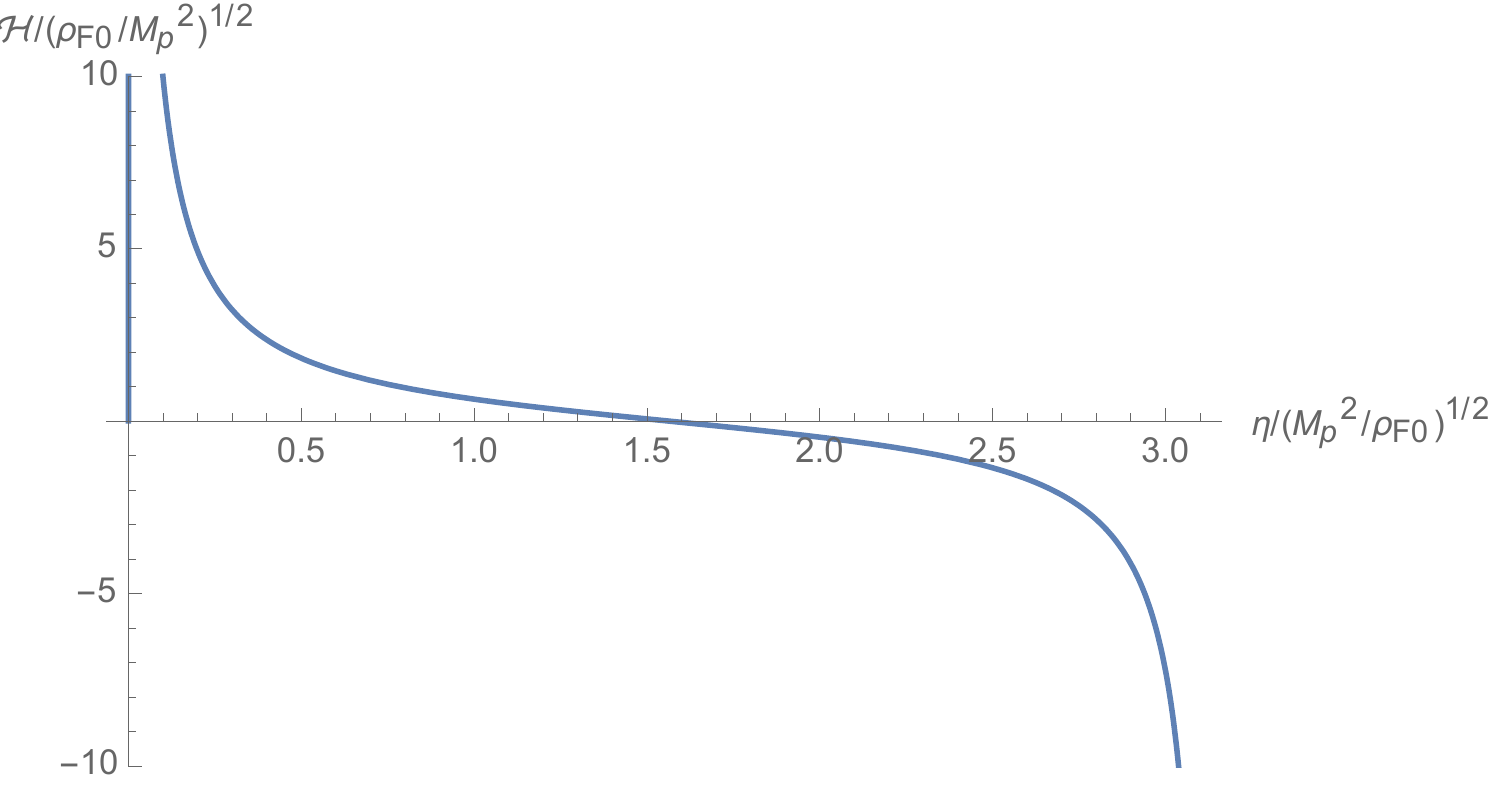}
\caption{\label{fig:hubble}Conformal Hubble rate, $\mathcal{H}=a'/a$, as a function of conformal time. $\eta$, for single cycle from one bounce to another. The initial ghost field velocity at the bounce is taken to be $\psi'(0)=10^3\rho_{F0}^{1/2}$, while $\Lambda^4=\rho_{F0}$ and $f=10^{-2}M_p$. In the first half of the evolution, $\mathcal{H}$ is positive and decreases as the universe expands. It reaches zero as the universe reaches its maximum extent and turns around. In the second half, $\mathcal{H}$ is negative and decreases as the universe contracts.}
\end{figure}

We can make some approximations to specialize these equations to the period of time near the bounce. Figure \ref{fig:hubble} shows an example of the evolution of the Hubble rate from one bounce to the next. The bounce occurs when the negative kinetic energy of the ghost field, $\psi$, becomes large enough to cancel the radiation energy density such that $\mathcal{H}=0$. At this point the potential term in equation \eqref{eq:psiEOM2} becomes irrelevant and we have approximately
\begin{equation}
\psi'' = -2\mathcal{H}\psi'.
\end{equation}
Using this in equation \eqref{eq:phi_k} and then using equation \eqref{eq:PertEE0i} we can get a decoupled equation for $\phi_k$,
\begin{equation}
\phi_k'' + 6\mathcal{H}\phi_k' + \left(k^2+2\mathcal{H}'+4\mathcal{H}^2-4\mathcal{K}\right)\phi_k = 0.
\end{equation}
Also, since during the bounce $\psi'$ is large, the cosine term in equation \eqref{eq:chi_k} oscillates quickly and averages to zero. We can therefore write the equation for $\chi_k$ as
\begin{equation}
\chi_k''+2\mathcal{H}\chi_k' +\left( k^2+\frac{2}{M_p^2}\psi'^2\right)\chi_k = -4\mathcal{H}\psi'\phi_k.
\end{equation}

\begin{figure}
\begin{subfigure}{0.47\textwidth}
\includegraphics[scale=0.8]{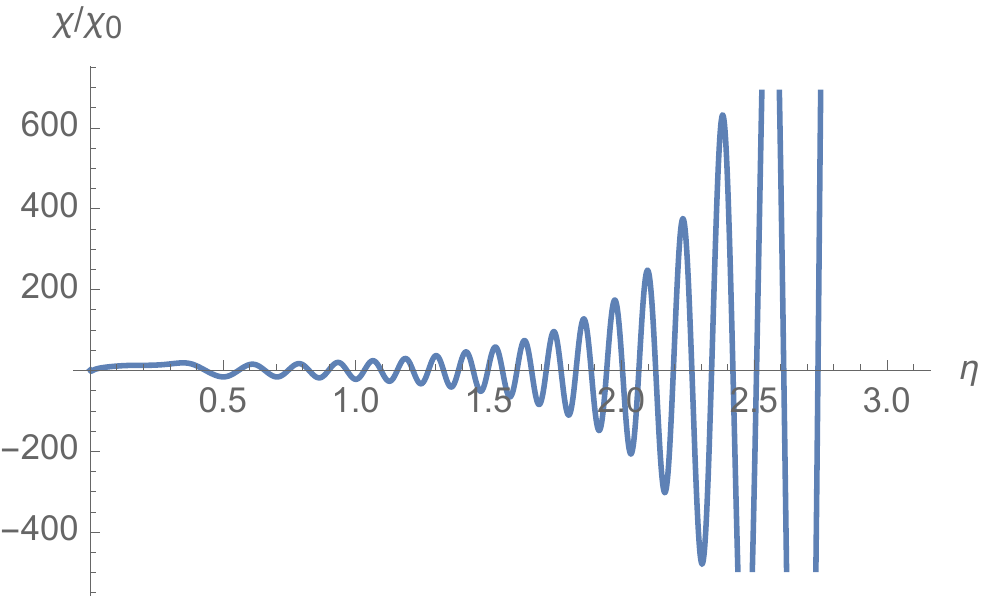}
\caption{$k=\sqrt{\rho_{F0}}/M_p$}
\end{subfigure}
\begin{subfigure}{0.47\textwidth}
\includegraphics[scale=0.8]{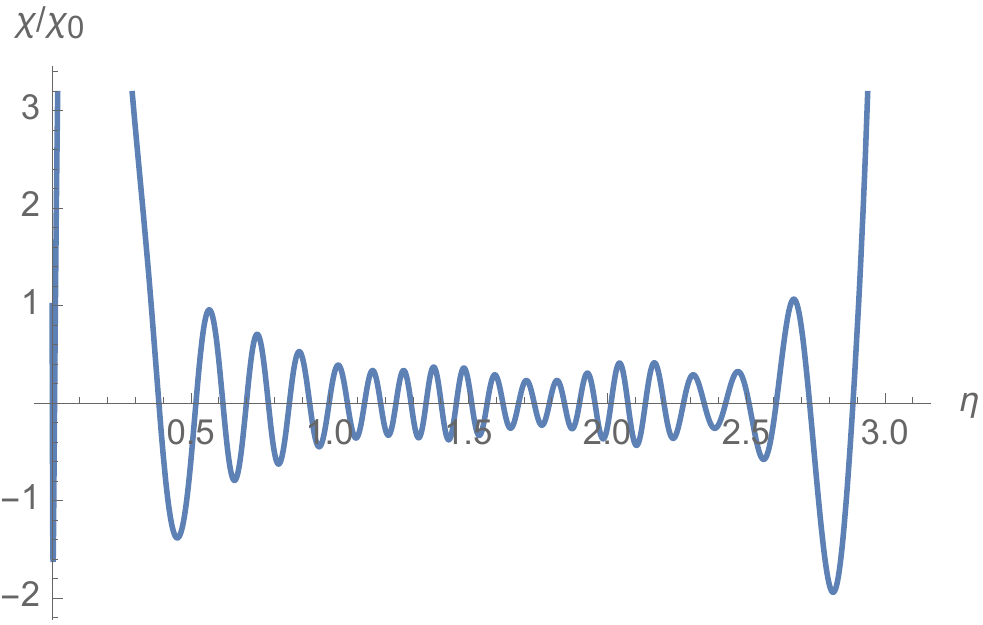}
\caption{$k=10\sqrt{\rho_{F0}}/M_p$}
\end{subfigure}
\begin{subfigure}{0.47\textwidth}
\includegraphics[scale=0.8]{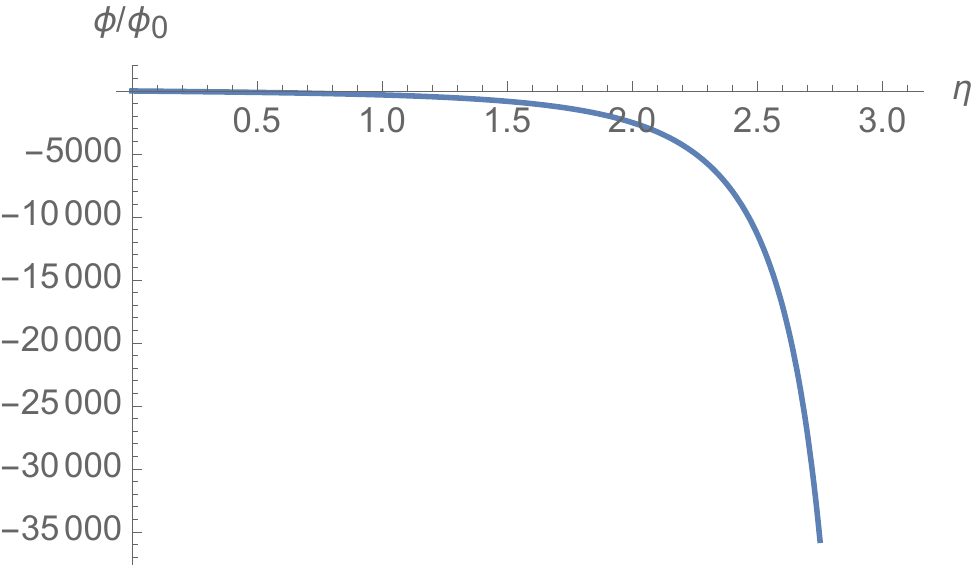}
\caption{$k=\sqrt{\rho_{F0}}/M_p$}
\end{subfigure}
\begin{subfigure}{0.47\textwidth}
\includegraphics[scale=0.8]{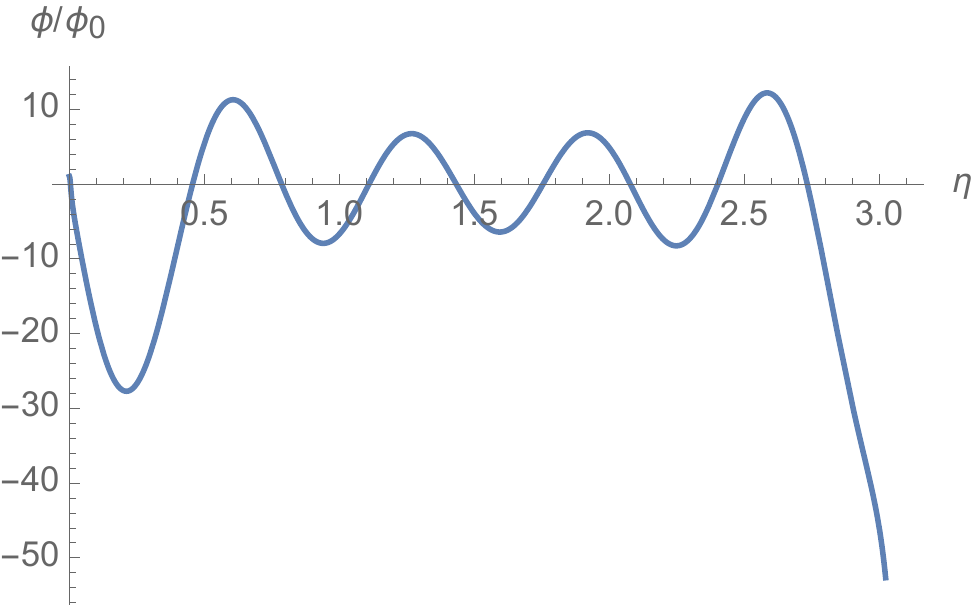}
\caption{$k=10\sqrt{\rho_{F0}}/M_p$}
\end{subfigure}
\caption{\label{fig:perts}Behaviour of the ghost field perturbation $\chi$, and the metric perturbation, $\phi$ during expansion phase. The parameters used are $\psi'(0)=10^3\rho_{F0}^{1/2}$, $\Lambda^4=\rho_{F0}$ and $f=10^{-2}M_p$. The conformal time, $\eta$ is in units of $(M_p^2/\rho_{F0})^{1/2}$. The amplitude of the perturbations are relative to their initial amplitudes at the bounce.}
\end{figure}

For the evolution of the perturbations during the expansion and contraction phases we use the full equations \eqref{eq:chi_k} and \eqref{eq:phi_k} along with the background equations \eqref{eq:psiEOM2}. In figure \ref{fig:perts} we show the behavior of the perturbations for two different $k$ modes. We can compare the size of the $k$ modes to the Hubble scale by referring to figure \ref{fig:hubble}. The perturbations which enter the horizon early, $k\gtrsim 10\sqrt{\rho_{F0}}/M_p$, are stable throughout the expansion and contraction of the universe. Those that enter the horizon later on in the expansion, $k\lesssim 10\sqrt{\rho_{F0}}/M_p$, are unstable. Linear, short wavelength modes of the ghost field are stable during the expansion and contraction phases. There are however classically unstable modes around the size of the Hubble scale during most of the expansion phase.

The main problem with the ghost field however is really the nonlinear instability due to the negative kinetic term, and its interaction with other fields.  This points us towards the idea of a ghost condensate to stabilize the field \cite{ArkaniHamed:2003uy}. The idea is to add higher order derivative terms which stabilize the ghost field around a non-zero value of $\partial_\mu\psi$, analogous to the stabilization of a tachyonic potential by adding higher order polynomials to the potential.

\section{Discussion}

In this work we studied the stability of a ghost field in the context of a cyclic universe scenario.  Unlike the linear instabilities that non-interacting ghosts suffer in flat, time independent backgrounds, we will argue that it is possible for ghosts in cyclic cosmologies to exhibit stability.  We already demonstrated that a background time-dependent ghost subject to an oscillatory potential exhibits limit cycles which bound the field trajectories and energies in a finite series of limit cycles.  This reflects the fact that the oscillatory potential bounds the negative energy background field configuration.  

Secondly we find that the gauge invariant coupled metric and ghost cosmological perturbations have a peculiar feature.  First, similar to well behaved scalar perturbations, the ghost field undergoes a classical instability for superhorizon modes.  However, subhorizon modes are well behaved and are generically oscillatory.  One distinct feature of the ghost system is that modes that are marginally sub-horizon are unstable and could actually have some interesting phenomenological consequences.   Since this class of sub horizon instabilities are classical, it could signal the formation of a stable condensed configuration.  Mukohyama have shown that ghost instabilities that accreted into black holes behave just like dust and could also serve as a viable dark matter candidate\cite{Muk}.   It is therefore possible the these unstable sub-horizon modes can accreted into primordial black-holes which accrete the ghost field and we are currently investigating this interesting possibility.

\begin{figure}
\begin{subfigure}{0.47\textwidth}
\includegraphics[scale=1]{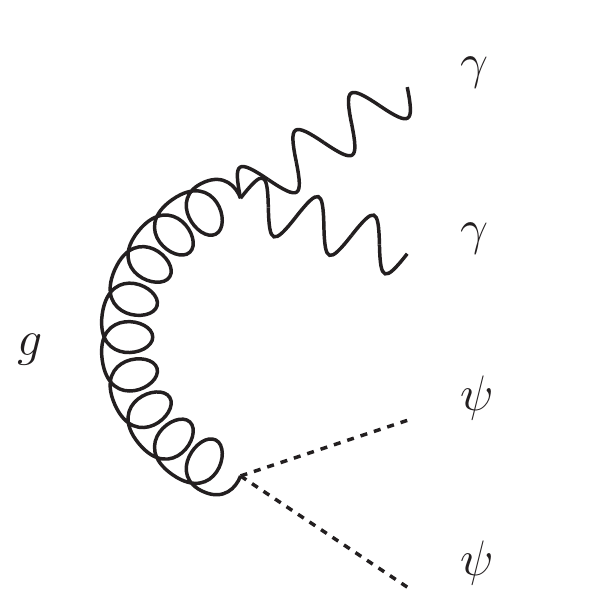}
\caption{\label{fig:gravdecay}Graviton mediated vacuum decay}
\end{subfigure}
\begin{subfigure}{0.47\textwidth}
\includegraphics[scale=1]{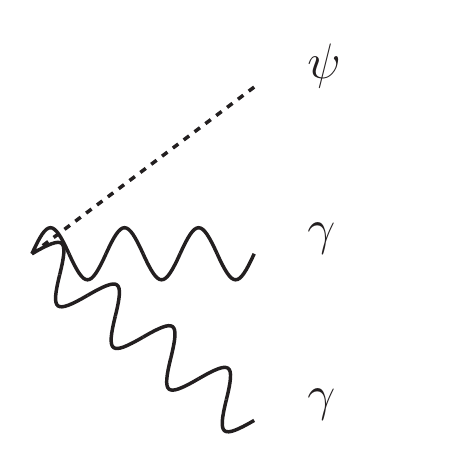}
\caption{\label{fig:photdecay}Direct coupling vacuum decay}
\end{subfigure}
\caption{Vacuum decay channels when ghost is present.}
\end{figure}

At the nonlinear level it is well established that the ghost vacuum can undergo graviton mediated decay into two photons and two ghosts via the process shown in figure \ref{fig:gravdecay}. The decay rate for this process is naively infinite, however, if we introduce a cutoff energy, $E_c$, above which the effective theory with the ghost does not apply, then we can estimate a decay rate of $\Gamma_{0\rightarrow 2\gamma \phi} \sim \frac{ E_c^{8}}{M_{p}^{4}} $ . Constraints come from diffuse gamma ray backgrounds to give a UV cutoff at $E_c \leq 3$ MeV \cite{Cline}.  Therefore, our theory will likewise be effective up to a MeV cutoff.   Moreover, the ghost also directly couples to the photon through the term $\Delta \mathcal{L}= -\frac{1}{4}e^{-2\psi/M_p}F^2$ providing another channel for vacuum decay. To lowest order we will have an interaction $(\psi/2M_p)F^2$ which allows the vacuum to decay to a ghost plus two photons as in figure \ref{fig:photdecay}. The interaction is Planck suppressed just as for the graviton mediated decay, however the process involves only a single vertex so the decay rate goes as $\Lambda^6/M_p^2$, an enhancement of $M_p^2/\Lambda^2$ over the graviton mediated channel.  This relatively small UV cutoff points to the need for a ghost free UV complete theory or another avenue to quantum mechanically stabilize the ghost system.  The cutoff induced by direct interactions of the ghost can be relaxed if the gauge field is identified as the dark photon.  When the ghost is directly coupled to regular photons, a bound can be found from high energy photon measurements.  However, in the case where the direct coupling is with the dark photon, the analogous constraint would also involve the suppressed coupled between the dark photon and standard model particles \cite{Essig:2013lka,An:2014twa,Ilten:2016tkc}.

One potential resolution to alleviating the issue of vacuum decay is to realize that the ghost condenses in the IR.  In this case, an opposite sign quartic kinetic self interaction with a frequency dispersion relation renders the vacuum stable against decay.   This idea is reminiscent to a Higgs-like phenomena of spontaneous symmetry breaking.  Here the Higgs field $\Phi$ has an unstable tachyonic mode around the false vacuum $\left<\Phi\right>=0$.   While fluctuations around field values are unstable, the full theory is stable since the Higgs potential has a global vacuum that is bounded from below.  Likewise, as pointed out in Arkani-Hamed et al. \cite{ArkaniHamed:2003uy}, the ghost field can be seen as an effective theory with higher order kinetic terms which has a stable minima:
\begin{equation}
S= \int d^{3}x dt\left[\frac{1}{2}M^{4}\dot{\phi}^{2} -\frac{1}{2}\tilde{M}(\nabla^{2}\phi)^{2} - \frac{M^{4}}{2c}\dot{\phi}(\nabla\phi)^{2} + \frac{M^{4}}{8c^{2}}(\nabla \phi)^{4} + ...\right]
\end{equation}

A power-counting analysis of scaling dimension of the above operator reveals that there are no large quantum instabilities in the IR.  Therefore if our ghost is a condensate in the IR we can evade the instabilities provided that there is a UV completion of our theory.
The issue of finding a UV completed theory that gives the ghost condensate is still an open ended quest.  Recently, the authors \cite{Krotov:2004if} of claimed that an Abelian-Higgs like model coupled to fermions can yield a ghost condensate after integrating out the fermions.  
%However it was later shown that the sign necessary for the quartic kinetic term depends on its regulator dependent \cite{O'Connell:2006de}.  
A roadblock to UV completion was pointed out in a very interesting work by Adams et. al where they demonstrated that the constraints of a local, Lorentz-invariant, S-matrix prohibit a UV completion of a ghost condensate further negating the claims of \cite{Adams:2006sv}.  The point is that a translationally invariant ghost configuration that picks a frame, necessarily generates superluminal waves which violate causality in the S-matrix.  However there is a very interesting loop-hole that was pointed out by Dvali and collaborators\cite{Dvali:2012zc}:

The road to UV completion rests on the Wilsonian paradigm which is based on the existence of weakly coupled degrees of freedom that become relevant in the deep UV.  A good example of this is in QCD where below some cutoff scale strongly coupled states such as pions arise.  Above the cutoff scale the pion is no longer a reliable degree of freedom and are replaced by weakly coupled quarks and gluons in the UV.  However, the authors of \cite{Dvali:2012zc} show another route, using ghost condensates as an example, where there is a non-Wilsonian completion.  In this case it is not new weakly coupled states that appear but collective excitations of multi-particle states composed out of soft original quanta.   Interestingly these field configurations are often non-linear and appear to be non-unitary.The authors show that the issue of superlumanility and non-unitarity can be resolved because even if the background classicalon solution produces super-luminal waves, boost transformations that can lead to acausality are not allowed by the background.  We trade a short spatial wavelength instability for a different instability due to higher order time derivatives, if we insist on local Lorentz invariance.  If we give up local Lorentz invariance then we run afoul of all experiments testing special relativity.  However, instabilities do not prove a theory is inconsistent, only that the vacuum and quasiparticles have not been properly identified.
%We will pursue the possibility as to whether our background ghost field falls into the category of a classicalon as a means to address the quantum stability of the cosmic ghost.

\end{document}